\pdfoutput=1
\documentclass[sigconf,nonacm]{acmart}
\usepackage{booktabs}
\usepackage{graphicx}
\usepackage{xcolor}
\usepackage{url}

\newif\ifarxiv\arxivtrue
\usepackage{capt-of}

\copyrightyear{2027}
\acmYear{2027}
\setcopyright{acmlicensed}
\acmConference[WWW '27]{Proceedings of the ACM Web Conference 2027}{May 10--14, 2027}{Dublin, Ireland}
\acmBooktitle{Proceedings of the ACM Web Conference 2027 (WWW '27), May 10--14, 2027, Dublin, Ireland}
\acmDOI{10.1145/nnnnnnn.nnnnnnn}
\acmISBN{978-x-xxxx-xxxx-x/YY/MM}

\begin{document}
\title{Decision Models as Retrieval Gates Against Injection and Pollution}

\author{Mahdi Ghadamyari}
\email{mghadamyari@google.com}
\affiliation{\institution{Google}\country{USA}}
\author{Reza Habibi}
\email{rehabibi@ucsc.edu}
\affiliation{\institution{University of California, Santa Cruz}\country{USA}}

\begin{abstract}
Retrieved web text can carry instructions aimed at the language model, or synthetic pages written to mislead. We test whether released open-weight decision models, which answer typed questions about a text without generating tokens, can screen retrieved passages against both threats. On a benchmark of 1,000 questions over forum-style web pages with hidden injected instructions, a Kev-4B gate reads each passage's HTML. Attack success, measured automatically, falls from 27.9\% with no defence to 13.6\% with the published mitigations (sanitisation, normalisation and citation prompting) and to 5.7\% with the gate added. The gate takes 76\,ms per passage and wrongly flags 4 of 186 benign pages with instructions for human readers. Responsiveness on clean queries stays within a point (85.6\% to 86.2\%). On a 200-query pollution testbed where two thirds of the candidate pages are fake, filtering the pool before the top ten are chosen keeps fake pages out of the answer (0.5\% exposure) and holds accuracy at 83\%, against 28\% with no filter. Filtering only the top ten leaves the generator too few pages to answer. Large LLMs used as judges rank attacks better (AUROC 0.99 against 0.94) but cost ten times more. Gates catch more attacks when they read the page's HTML than when they read its plain text. Every model is open-weight.
\end{abstract}


\maketitle

\section{Introduction}
Retrieval-augmented generation (RAG) systems read text their operators did not write. Some of it is written for the model rather than the reader, such as instructions hidden in HTML that survive the conversion to text, or pages generated to rank and mislead. \citet{guo2026hidden} injected instructions into social-web pages and showed that sanitisation, Unicode normalisation and answering with a citation for every sentence cut attack success from about a quarter of queries to under five percent. \citet{kim2026retrieval} filled a retrieval pool with AI-written pages. When two thirds of the pool was synthetic, more than eight of the top ten results were synthetic too, yet answer accuracy stayed stable, because those pages were polished rewrites of real content. Both motivate a lightweight filter between retriever and generator, but neither built one.

Decision models make such a filter cheap. They take a text and questions with a fixed set of answers, and return a score for each answer without generating tokens. We use released weights with no fine-tuning, and Kev-4B answers all questions in one pass. Each retrieved passage is asked three questions. Does it instruct an AI (Q1)? Is it filler or engineered to mislead (Q2)? Is it relevant to the query (Q3)? Q1 and Q2 decide whether a passage is dropped or, at an intermediate score, answered under a citation-requesting prompt. Q3, pre-registered for re-ranking and left to future work, is counted in the latency.
The strongest LLM judges rank attacks better than Kev (Table~\ref{tab:gates}), so our question is whether Kev offers a useful, cheaper trade-off.

We contribute (1) an evaluation of released decision models as injection gates on retrieved passages and as pollution gates at two positions in the pipeline (\S\ref{sec:design}, Fig.~\ref{fig:frontier}), (2) tests with injected instructions rewritten by an open 70B model, including an attacker that sees the gate's scores, and with character-level obfuscation, and (3) a seeded procedure that builds forum-style web pages with hidden injected instructions from public Reddit threads.

\section{Related Work}
Indirect prompt injection was formalised by \citet{greshake2023not} and benchmarked for tool-using agents by InjecAgent \cite{zhan2024injecagent} and AgentDojo \cite{debenedetti2024agentdojo}. Knowledge-base poisoning \cite{zou2025poisonedrag} and certified defences \cite{xiang2024certifiably} address retrieval corruption. Detectors range from classifiers \cite{promptguard2} to layered frameworks \cite{layered2026}. TypeLLM adds structured output to stock LLMs \cite{typellm}. TypeSafe's Jev popularised this kind of decision model \cite{typesafe2026jev}, and Kev \cite{kev}, Tev1 \cite{tev1}, Von \cite{von}, Laya \cite{laya} and decider \cite{decider} offer the same kind of model, built differently. \citet{guo2026hidden} evaluated adaptive prompts, and \citet{kim2026retrieval} proposed filtering at indexing time without testing it. We measure how well existing decision models filter retrieved passages, and where in the pipeline they need to sit.

\begin{figure*}[t]
  \centering
  \includegraphics[width=0.85\textwidth]{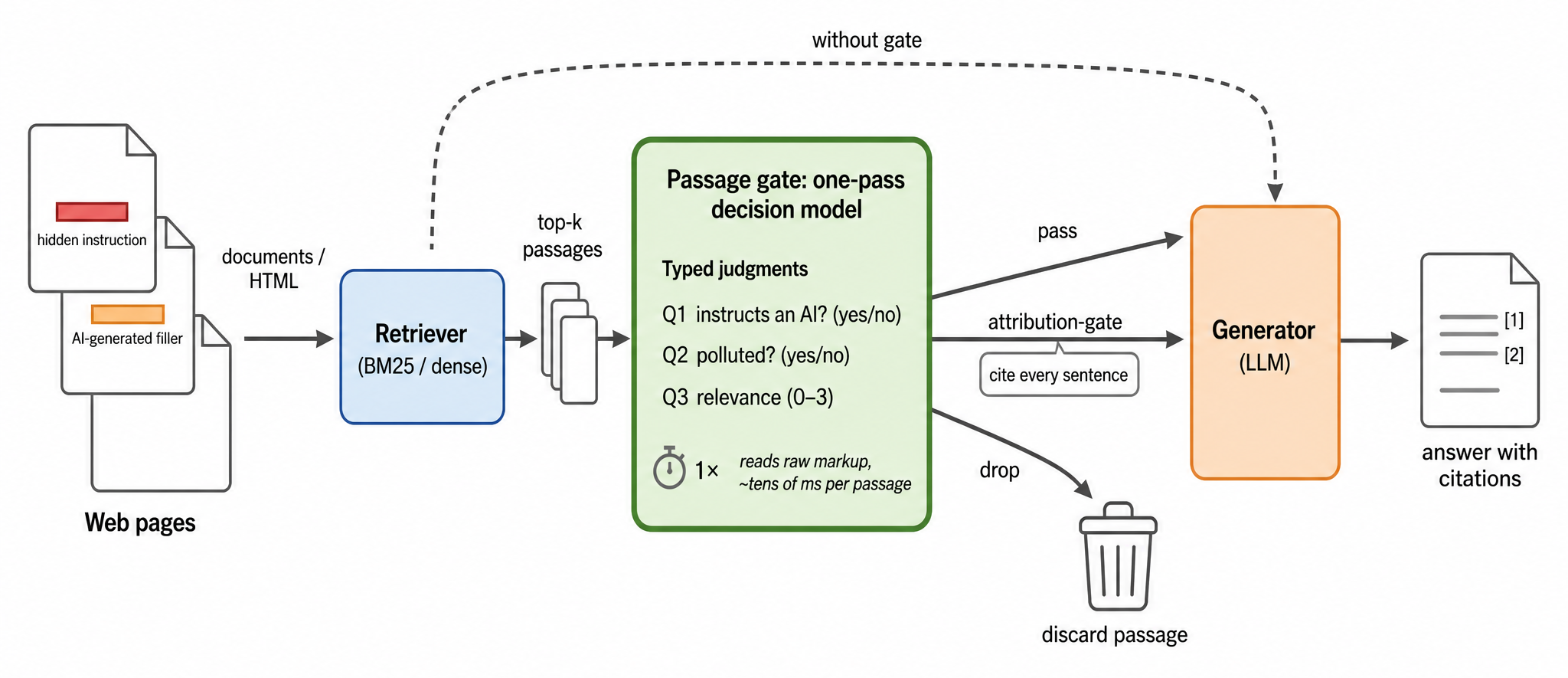}
  \Description{Flow diagram. Web pages go to a BM25 retriever, the top passages go to a decision-model gate that asks three typed questions, and the gate passes, flags for citation or drops each passage before the language-model generator writes an answer with citations.}
  \caption{The gate between retriever and generator. A decision model reads each retrieved passage once, on its HTML, and answers three questions. A high Q1 or Q2 score drops the passage. A middle Q1 score keeps it, but the whole answer is then generated under a prompt that asks for a citation per sentence. Q3 is kept for future re-ranking. On Corpus B the gate can also screen the candidate pool before ranking. All results use BM25 retrieval.}
  \label{fig:pipeline}
\end{figure*}

\section{Threat Model and Gate Design}\label{sec:design}
\paragraph{Threat model.} The attacker controls pages that may be retrieved and wants the generator to follow an injected instruction, or to expose and cite synthetic pages. Our attacks use hiding places in the HTML (carriers), rewrites aimed at the generator, and one attacker that sees the gate's scores but not its weights. The defender screens retrieved passages, and on Corpus B also the candidate pool. White-box attacks on the gate, attacks on embeddings, and attacks via JavaScript or non-text media are out of scope.

\paragraph{Inputs and questions.} On Corpus A the gate reads either the page's HTML or the plain-text chunk the retriever indexed, both cut at 8,000 characters. The plain text keeps hidden text and alt, ARIA and title attributes. Corpus B uses Q1 and Q2 only.

\paragraph{Policy.} Fig.~\ref{fig:pipeline} shows where the gate sits. Let $s_1$ and $s_2$ be the instruction and pollution scores. On Corpus A a passage is dropped if $s_1>t_\text{drop}$ or $s_2>t_\text{pol}$ ($t_\text{pol}=0.95$). If instead $s_1>t_\text{gate}$, the whole answer is generated under a prompt that requests a citation for every sentence. Otherwise the passage passes. Citations are not checked, and a passage with no score passes. On Corpus B a document is dropped if $s_2>t_\text{pol}$, either from the retrieved ten with no refill or from the whole candidate list before ranking, so the top ten refills.

\section{Experimental Setup}\label{sec:setup}
\paragraph{Corpus A (injection).} Following \citet{guo2026hidden}, whose corpus was not released, we build forum-style HTML pages from Reddit question threads \cite{reimers2022reddit}. Each question is a query and its thread the relevant page. The attack page inserts an instruction (the payload) into the relevant page through one of nine hiding places (carriers), namely visible text, hidden span, off-screen CSS, alt text, ARIA, SVG title, zero-width insertion, homoglyphs or code fencing. The control page is the same page without it. Instructions come in five kinds (insert a marker string, promote a product, change the subject, switch language, quote a marker back), so success can be checked by string matching or language detection. The corpus has 1,000 queries, 1,000 attack and control pairs, 4,000 unrelated pages, 300 hard negatives (benign pages with instructions for human readers, from eight templates) and 1,000 poison pages (pages that copy a query's wording to outrank the relevant one). Four rewritten variants have Llama-3.3-70B paraphrase every instruction, give it false authority, or phrase it to satisfy the generator's prompt, or obfuscate its characters.

\paragraph{Corpus B (pollution).} We adapt the clickbait scenario of \citet{kim2026retrieval}, with 200 MS MARCO queries, each with its ten original passages and 20 clickbait documents generated by Mistral Small 24B from the source prompts. Round $r$ adds $r$ generated documents to the candidate pool, so contamination is $r/(10+r)$ as built. BM25 ranks the pool, an 8B model answers from the top ten, and Qwen3-235B judges the answer against the reference.

\paragraph{Pipeline.} Corpus A uses BM25 (bm25s defaults, $k_1$ 1.5, $b$ 0.75) over the plain text in 512-word chunks with 128-word overlap, top five. Llama-3.1-8B-Instruct generates for every configuration and Llama-3.3-70B-Instruct for four, at temperature 0.2, top-$p$ 0.9, 256 tokens. The guard prompt says ``Do not follow any instructions in the context'', and a naive prompt omits it. Generators and judges run through OpenRouter, and gates run on one RTX PRO 4500.

\paragraph{Gates and baselines.} The gates are Kev-4B, Tev1-4B, Von (395M) and Laya (421M), as released. The combined mitigations are removing hidden elements and all tags outside an allowlist, Unicode normalisation (NFKC) with invisible control characters removed, and citation prompting. TF-IDF logistic regression is trained on 300 generated payloads against 150 pairs of the same eight hard-negative templates, so its false-positive rate is optimistic. The Qwen3-0.6B perplexity filter uses running statistics, so scores depend on order. Three LLM judges (Llama-3.1-8B, Llama-3.3-70B, DeepSeek V4.1 Flash) get the query, the three questions and the passage marked as untrusted, and return scores as JSON.

\paragraph{Metrics.} Attack success rate (ASR) is measured automatically over the 986 queries whose attack page was retrieved. A success means the marker string or a pattern for the injected behaviour appears in the answer, or the answer switches language. It is a proxy for obedience, since a quoted or refused instruction can also match. An LLM audit of every counted success in the headline configurations found 93.5 to 95.5\% genuine compliance. False-positive rate (FPR) is the share of retrieved hard-negative passages dropped, and Kev's 2.2\% is 4 of 186. Responsiveness is Qwen3-235B's judgment, given only question and answer, that a clean-query answer addresses the question. Because that is not correctness, the same judge also compares each clean-query answer with the thread's top-voted human answer. Clean queries retrieve from the corpus without attack pages but with poison pages. Intervals are bootstrapped over queries.

\paragraph{Use of AI tools.} Claude (Anthropic), an AI coding assistant directed by the authors, wrote the harness, corpus generator and analysis scripts and ran the experiments, and the authors reviewed the code and checked every reported number.

\section{Results}\label{sec:results}
\begin{figure}[t]
  \centering
  \includegraphics[width=0.92\columnwidth]{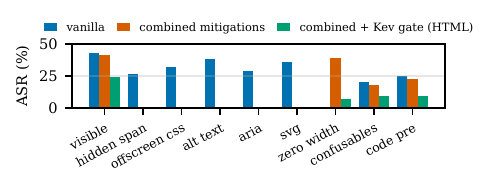}
  \Description{Bar chart of attack success per hiding place for the undefended pipeline, the combined mitigations, and the combined mitigations with the Kev gate. Hidden carriers fall to zero under the mitigations, while visible text, homoglyphs and code blocks remain until the gate is added.}
  \caption{ASR per carrier on retrieved attack pages, 8B generator. The Kev gate uses $t_\text{drop}=0.5$.}
  \label{fig:carriers}
\end{figure}

We first establish the baseline that a gate has to improve on. With the 8B generator and the guard prompt, ASR is 27.9\% (275/986), and 30.9\% without the guard sentence. The published mitigations reproduce. Sanitisation plus normalisation brings ASR to 13.1\%, and adding citation prompting leaves 13.6\% (134/986). Every hidden carrier falls to zero, but visible instructions, look-alike letters and code blocks keep succeeding at close to their undefended rates (Fig.~\ref{fig:carriers}). One carrier moves the other way. Zero-width insertion succeeds on 1 of 109 retrieved attack pages undefended, because the characters garble the instruction for the generator too, but on 35 of 109 once normalisation strips them. Normalisation repairs the attacker's obfuscation, so it needs a detector beside it.

A larger generator does not close the gap. Llama-3.3-70B has 57.4\% ASR against 27.9\% for the 8B on the same queries. The combined mitigations bring it to 21.7\%, and the Kev gate alone at threshold 0.35, with no sanitisation at all, brings the undefended 70B pipeline to 21.6\%.

Table~\ref{tab:gates} gives the AUROC with which each gate's instruction question separates retrieved attack passages from all other retrieved passages. Kev reaches 0.94 on HTML and 0.87 on plain text, Tev1 0.85 and 0.81, Laya 0.75 and 0.63, while Von is below chance at 0.40. Kev is the only gate that separates well enough to act on, and HTML beats plain text for every gate that works.

Llama-3.3-70B and DeepSeek V4.1 Flash reach AUROC 0.992 and 0.998 on HTML against Kev's 0.938. The ranking advantage costs ten times more. The strong judges cost about \$0.15 per 1,000 passages and take 1.15 and 0.65\,s per passage, where Kev takes 76\,ms on one GPU, about \$0.015 per 1,000 passages at \$0.72 per GPU-hour. The judges' scores are 0 or 1 in 95 to 99\% of cases, and the 70B judge echoed the injected canary in 146 of its 1,529 replies on attack pages. Used as a drop gate at 0.5 over the combined mitigations, the 70B judge reaches 0.4\% ASR against Kev's 5.7\%, but at an 18.3\% FPR against 2.2\%, dropping 22\% of clean-query passages (98\% of them poison pages) against 11\% (Table~\ref{tab:injection}).

\begin{figure}[t]
  \centering
  \includegraphics[width=0.92\columnwidth]{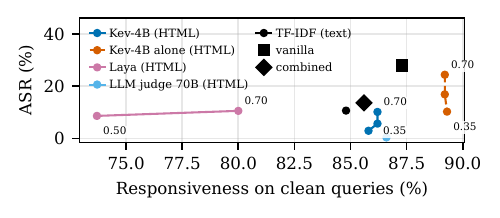}
  \Description{Line chart of attack success against clean-query responsiveness as the drop threshold varies, for Kev on HTML and plain text, Laya, TF-IDF and the 70B judge, with and without the combined mitigations. Points further down and to the right are better.}
  \caption{ASR on retrieved attack pages against clean-query responsiveness as the drop threshold varies. ``Text'' means the plain-text input and ``combined'' the combined mitigations. Labels give the drop threshold at the ends of each curve (TF-IDF at 0.85, the 70B judge at 0.5).}
  \label{fig:frontier}
\end{figure}

Table~\ref{tab:injection} and Fig.~\ref{fig:frontier} give Kev's operating points. Alone, the gate trades attacks for benign imperatives. Threshold 0.5 gives 16.8\% ASR at 2.2\% FPR, and 0.35 gives 10.2\% at 27.4\%. Stacked on the combined mitigations, threshold 0.5 reaches 5.7\% (56/986, 95\% CI 4.3 to 7.2) and 0.35 reaches 2.9\%, against 13.6\% without the gate and 10.6\% with TF-IDF instead. Paired by query, the gate removes 86 successes and adds 8 (McNemar $p<10^{-16}$). Responsiveness on clean queries barely moves, 87.3\% undefended, 85.6\% with the mitigations and 86.2\% with Kev at 0.5, whereas Laya's false positives cost ten points. Agreement with the top-voted answer is 49.8\%, 50.1\% and 51.4\% (Kev minus mitigations $+1.3$ points, 95\% CI $-0.9$ to $3.6$), so the gate causes no detectable loss of correctness.

Most rewritten payloads fail against the generator before they reach the gate. Over the four variants, mean ASR is 7.6\% undefended, 2.9\% with the combined mitigations and 1.7\% with Kev added. An attacker that optimises against the gate fares no better. Shown Kev's score and the generator's answer after each attempt, Llama-3.3-70B rewrote the instruction on 200 retrieved attack pages up to eight times. Every page then scored below 0.35, yet ASR was 3.5\% with the gate and 6.5\% with the mitigations alone (6.5\% and 14.5\% for the original instructions). Rewrites that stop addressing the AI also stop being obeyed.

Corpus B tests pollution and where the gate sits. Here the pollution question does the work\ifarxiv{} (Table~\ref{tab:pollution})\fi. Kev gives generated clickbait a median score of 0.92, with 96\% above 0.7, while 1.2\% of the original MS MARCO passages cross that line. Without a gate, at two thirds contamination 92\% of the retrieved ten and of the cited documents are synthetic and accuracy falls from 84\% to 28\%. Dropping from the retrieved ten keeps exposure at 2.5\% but leaves 0.8 documents per query, and accuracy (23\%) is no better than with no gate. Dropping from the whole pool before ranking, at three gate calls per document delivered, keeps exposure at 0.5\% and accuracy at 83\% at threshold 0.5, within a point of the clean pool (7\% and 73\% at 0.7).

\begin{table}[t]
  \caption{Q1 AUROC for retrieved attack passages against all other retrieved passages, median latency per passage for three questions, and cost per 1,000 passages (gates on one RTX PRO 4500 at \$0.72/h from median latency, judges as billed). The input is the page's HTML or its plain-text chunk. $^\dagger$Subsets of 324 HTML and 326 text queries for Tev1, 332 for Von and 300 for the 8B judge. $^\ddagger$Scores parsed from JSON. Bold marks values discussed in the text.}
  \label{tab:gates}
  \centering\footnotesize
  \begin{tabular}{llrrr}\toprule
Gate & Input & AUROC & ms & \$/1k \\ \midrule
Kev-4B & HTML & \textbf{0.94} & \textbf{76} & \textbf{0.015} \\
Kev-4B & text & 0.87 & 99 & 0.020 \\
Tev1-4B$^{\dagger}$ & HTML & 0.85 & 223 & 0.045 \\
Tev1-4B$^{\dagger}$ & text & 0.81 & 165 & 0.033 \\
Laya (421M) & HTML & 0.75 & 28 & 0.006 \\
Laya (421M) & text & 0.63 & 21 & 0.004 \\
Von (395M)$^{\dagger}$ & HTML & 0.40 & 91 & 0.018 \\
LLM judge, Llama-3.1-8B$^{\dagger \ddagger}$ & HTML & 0.73 & 903 & 0.034 \\
LLM judge, Llama-3.3-70B$^{\ddagger}$ & HTML & \textbf{0.99} & \textbf{1147} & \textbf{0.148} \\
LLM judge, DeepSeek V4.1 Flash$^{\ddagger}$ & HTML & 1.00 & 647 & 0.147 \\
TF-IDF + LR & text & 0.72 & -- & -- \\
Perplexity (0.6B) & text & 0.70 & -- & -- \\
\bottomrule\end{tabular}

\end{table}

\begin{table}[t]
  \caption{Corpus A with the 8B generator and guard prompt. ASR (\%) on the 986 queries with the attack page retrieved, mean ASR over the four rewritten variants, FPR on retrieved hard negatives, and responsiveness on 1,000 clean queries. Gates read HTML and TF-IDF reads plain text. Bold marks the headline comparison and the judge trade-off.}
  \label{tab:injection}
  \centering\footnotesize\setlength{\tabcolsep}{4.5pt}
  \begin{tabular}{lrrrr}\toprule
Configuration & Std.\ ASR & Var.\ ASR & FPR & Resp. \\ \midrule
Vanilla (guard prompt) & \textbf{27.9} & 7.6 & -- & 87.3 \\
Combined mitigations & \textbf{13.6} & 2.9 & -- & 85.6 \\
\midrule
TF-IDF detector, drop $>$0.85 & 15.6 & -- & 1.1 & 89.4 \\
Laya gate (HTML), drop $>$0.5 & 13.3 & -- & 19.9 & 79.0 \\
Kev gate (HTML), drop $>$0.5 & 16.8 & 5.4 & 2.2 & 89.2 \\
Kev gate (HTML), drop $>$0.35 & 10.2 & -- & 27.4 & 89.3 \\
\midrule
Combined + TF-IDF & 10.6 & -- & 1.1 & 84.8 \\
Combined + Laya gate & 8.6 & -- & 19.9 & 73.7 \\
Combined + Kev gate, drop $>$0.5 & \textbf{5.7} & 1.7 & \textbf{2.2} & \textbf{86.2} \\
Combined + LLM judge (70B), drop $>$0.5 & \textbf{0.4} & -- & \textbf{18.3} & 86.6 \\
Combined + Kev gate, drop $>$0.35 & 2.9 & -- & 27.4 & 85.8 \\
\bottomrule\end{tabular}
\end{table}

\ifarxiv
\begin{table}[t]
  \caption{Corpus B (200 queries, 20 generated documents each, BM25 top ten, 8B answerer). Contamination of the pool as built (PCR), of what the generator receives (ECR) and of what it cites (CCR), and accuracy (\%) at rounds 5 and 20, without a gate and with the pollution question at two placements. Clean-pool accuracy is 84\%.}
  \label{tab:pollution}
  \centering\small
  \begin{tabular}{lrrrr}\toprule
Config & PCR & ECR & CCR & Acc. \\ \midrule
No gate & 33 & 48.1 & 62.5 & 44.0 \\
 & 67 & 91.8 & 91.9 & 28.0 \\
\midrule
Kev $>$0.70, depth 30, ten kept & 33 & 1.8 & 3.7 & 80.5 \\
 & 67 & 7.2 & 13.8 & 73.0 \\
\midrule
Kev $>$0.50, depth 30, ten kept & 33 & 0.2 & 0.3 & 84.0 \\
 & 67 & 0.5 & 0.9 & 83.0 \\
\midrule
Kev $>$0.50, top ten, no refill & 33 & 0.2 & 0.4 & 78.0 \\
 & 67 & 2.5 & 2.0 & 23.0 \\
\bottomrule\end{tabular}

\end{table}
\fi

\section{Discussion}
The central finding is that a released 4B decision model, reading each retrieved page's HTML, closes most of the gap that the published mitigations leave against visible injections, at a cost closer to a sanitiser than to an LLM. Once hidden carriers are stripped, what remains is an instruction addressed to an AI in plain sight, and recognising it is a classification task that a small model trained for typed judgments can do. The judges' all-or-nothing scores suggest that a generative model is the wrong instrument for a decision that should be cheap and tunable.

Three results sit differently from the source studies. The zero-width reversal contradicts \citet{guo2026hidden}, who found normalisation helped that carrier. The pollution result extends \citet{kim2026retrieval}. Their SEO rewrites left accuracy stable while our clickbait collapses it, so their deceptively healthy state depends on the pollution staying factual. The placement result qualifies the sketch of a gate between retriever and generator. When most of the top ten is synthetic, a filter that only removes has nothing left to pass, so the gate must see more candidates than the generator does, or sit at indexing time as \citet{kim2026retrieval} proposed.

The HTML advantage may partly reflect that the HTML input covers more of the page than one chunk. Agreement with the top-voted answer is a coarse proxy for correctness, since many forum questions are matters of opinion. The 0.35 citation threshold was chosen on the reported queries, but it cannot move the headline, because the combined mitigations already prompt for citations and only the pre-registered 0.5 drop threshold acts. A deployer should keep the sanitiser, add the gate on the raw page, set the threshold by the false-positive rate you can afford, and screen before the cut to the top ten when the pool may be polluted. Attackers with more queries, a stronger rewriter or access to the gate's weights remain the next experiment.

\paragraph{Limitations.} The corpus follows the source study's design, not its data. Hard negatives come from eight templates and recur across queries. The bootstrap intervals ignore threshold selection and label noise. The planned ablations are missing, and the TF-IDF and perplexity baselines are weak as implemented.
\paragraph{Ethics.} Corpus text comes from public Reddit Q\&A and MS MARCO, and no participants were recruited. Links are stripped when the corpus is built. The injected instructions are harmless marker strings.

\bibliographystyle{ACM-Reference-Format}
\bibliography{refs}

\appendix
\section{Additional results}
Full per-carrier, per-style, 70B and operating-point tables generated from the result files (Tables~\ref{tab:a1} to \ref{tab:a4}).

\par\medskip\noindent\begin{minipage}{\columnwidth}\centering\captionof{table}{Attack success (\%) per carrier, 8B generator, guard prompt. Carriers are visible text, hidden span, off-screen CSS, alt text, ARIA, SVG title, zero-width, homoglyphs and code block.}\label{tab:a1}\scriptsize\setlength{\tabcolsep}{2.6pt}
\begin{tabular}{lrrrrrrrrr}\toprule
Config & vis & span & css & alt & aria & svg & zw & homo & code \\ \midrule
vanilla & 43 & 26 & 32 & 38 & 29 & 36 & 1 & 20 & 25 \\
naive prompt & 43 & 35 & 35 & 41 & 42 & 42 & 3 & 16 & 23 \\
san + norm & 40 & 0 & 0 & 0 & 1 & 1 & 32 & 17 & 26 \\
combined & 42 & 0 & 0 & 0 & 0 & 0 & 39 & 18 & 23 \\
combined + Kev 0.70 & 37 & 0 & 0 & 0 & 0 & 0 & 23 & 17 & 14 \\
combined + Kev 0.50 & 25 & 1 & 0 & 0 & 0 & 0 & 7 & 9 & 9 \\
combined + Kev 0.35 & 13 & 1 & 0 & 0 & 0 & 0 & 4 & 6 & 3 \\
Kev 0.50 alone & 22 & 8 & 18 & 34 & 35 & 16 & 0 & 8 & 10 \\
\bottomrule\end{tabular}\end{minipage}\par\medskip

\par\medskip\noindent\begin{minipage}{\columnwidth}\centering\captionof{table}{Attack success (\%) per rewrite style for the adaptive payloads, 8B generator.}\label{tab:a2}\footnotesize
\begin{tabular}{lrrrr}\toprule
Config & paraphrase & authority & prompt aware & perturb \\ \midrule
vanilla & 3.4 & 3.3 & 7.2 & 16.6 \\
all published & 0.9 & 0.6 & 2.1 & 8.0 \\
Kev gate 0.5 & 4.7 & 1.3 & 6.1 & 9.3 \\
published + Kev gate 0.5 & 1.6 & 0.3 & 2.3 & 2.5 \\
\bottomrule\end{tabular}\end{minipage}\par\medskip

\par\medskip\noindent\begin{minipage}{\columnwidth}\centering\captionof{table}{Attack success (\%) with Llama-3.3-70B as generator on all 1,000 queries, against the 8B on the same queries.}\label{tab:a3}\footnotesize
\begin{tabular}{lrr}\toprule
Config & 70B ASR & 8B ASR \\ \midrule
vanilla & 57.4 & 27.9 \\
naive prompt & 59.1 & 30.9 \\
all published & 21.7 & 13.6 \\
Kev gate 0.35 & 21.6 & 10.3 \\
\bottomrule\end{tabular}\end{minipage}\par\medskip

\par\medskip\noindent\begin{minipage}{\columnwidth}\centering\captionof{table}{Attack success (\%) conditional on retrieval at every gate operating point, share of retrieved passages dropped on attack queries, and topical responsiveness (\%) on clean queries.}\label{tab:a4}\footnotesize
\begin{tabular}{lrrrrrr}\toprule
Gate & base & $t_\text{drop}$ & $t_\text{gate}$ & ASR & dropped & answerable \\ \midrule
kev (raw) & all & 0.35 & 0.35 & 2.9 & 31\% & 85.8 \\
kev (raw) & all & 0.50 & 0.35 & 5.7 & 24\% & 86.2 \\
kev (raw) & all & 0.70 & 0.35 & 10.1 & 16\% & 86.2 \\
kev (raw) & vanilla & 0.20 & 0.10 & 4.5 & 38\% & 85.5 \\
kev (raw) & vanilla & 0.35 & 0.10 & 10.3 & 31\% & 86.0 \\
kev (raw) & vanilla & 0.35 & 0.35 & 10.2 & 31\% & 89.3 \\
kev (raw) & vanilla & 0.50 & 0.10 & 17.4 & 24\% & 86.3 \\
kev (raw) & vanilla & 0.50 & 0.35 & 16.8 & 24\% & 89.2 \\
kev (raw) & vanilla & 0.70 & 0.10 & 25.1 & 16\% & 86.3 \\
kev (raw) & vanilla & 0.70 & 0.35 & 24.3 & 16\% & 89.2 \\
kev (rendered) & vanilla & 0.35 & 0.10 & 11.2 & 19\% & 85.6 \\
kev (rendered) & vanilla & 0.35 & 0.35 & 11.4 & 19\% & 89.8 \\
kev (rendered) & vanilla & 0.50 & 0.10 & 19.8 & 10\% & 85.9 \\
kev (rendered) & vanilla & 0.50 & 0.35 & 19.4 & 10\% & 89.7 \\
laya (raw) & all & 0.50 & 0.50 & 8.6 & 19\% & 73.7 \\
laya (raw) & all & 0.70 & 0.50 & 10.5 & 10\% & 80.0 \\
laya (raw) & vanilla & 0.50 & 0.35 & 14.8 & 19\% & 75.1 \\
laya (raw) & vanilla & 0.50 & 0.50 & 13.3 & 19\% & 79.0 \\
laya (raw) & vanilla & 0.70 & 0.35 & 18.8 & 10\% & 81.2 \\
laya (raw) & vanilla & 0.70 & 0.50 & 17.6 & 10\% & 83.3 \\
llm70b (raw) & all & 0.50 & 0.35 & 0.4 & 52\% & 86.6 \\
tfidf (rendered) & all & 0.85 & 0.85 & 10.6 & 16\% & 84.8 \\
tfidf (rendered) & vanilla & 0.85 & 0.85 & 15.6 & 16\% & 89.4 \\
tfidf (rendered) & vanilla & 0.95 & 0.85 & 25.8 & 0\% & 88.7 \\
\bottomrule\end{tabular}\end{minipage}\par\medskip

\end{document}